\documentclass[12pt]{article}
\input{q-fou-tansform.sty}
\input{MyQcircuit.sty}

\begin{document}

\title{
Quantum Fast Fourier Transform\\
Viewed as a Special Case of\\
Recursive Application of\\
Cosine-Sine Decomposition}

\author{Robert R. Tucci\\
        P.O. Box 226\\
        Bedford,  MA   01730\\
        tucci@ar-tiste.com}

\date{ \today}

\maketitle

\vskip2cm
\section*{Abstract}
A quantum compiler is  a
software program for
decomposing (``compiling") an
arbitrary unitary matrix into
a sequence of elementary operations (SEO).
Coppersmith showed that the
$\nb$-bit Discrete Fourier Transform
matrix $U_{FT}$ can be decomposed in a very
efficient way,
as a sequence of
order($\nb^2$) elementary operations.
Can a quantum compiler
that doesn't know a priori
about Coppersmith's decomposition
nevertheless decompose
$U_{FT}$ as a
sequence of
order($\nb^2$) elementary operations?
In other words, can it rediscover
Coppersmith's decomposition
by following a much more general algorithm? Yes
it can,
if that more general algorithm is
the recursive application
of the Cosine-Sine Decomposition (CSD).

\newpage
\section{Introduction}

In quantum computing,
elementary operations are operations that
act on only a few (usually
one or two) qubits. For example, CNOTs and
one-qubit rotations are elementary operations.
A quantum compiling algorithm
is an algorithm for
decomposing (``compiling") an
arbitrary unitary matrix into
a sequence of elementary operations (SEO).
A quantum compiler is a software program
that implements a quantum compiling algorithm.

Henceforth, we will refer to
Ref.\cite{Tuc99} as Tuc99.
Tuc99 gives a quantum compiling algorithm,
implemented in a software program called
Qubiter.
The Tuc99 algorithm uses
a matrix decomposition called
the Cosine-Sine Decomposition (CSD)
which is well known
in the field of
Computational Linear Algebra\cite{Golub}.
Tuc99 uses CSD
in a recursive manner.
Henceforth we will refer
to the recursive application of
CSD as re-CSD or reap-CSD.

A modest desideratum for a
quantum compiler is that it should
recognize when  a matrix is
a tensor product of one-bit operators,
and decompose such a matrix
into a tensor product of one-bit
operators.
Qubiter does this
for the
$\nb$-bit Hadamard matrix $H_\nb$.

In Ref.\cite{Copper}, Coppersmith
showed how to express the
$\nb$-bit Discrete Fourier Transform
matrix $U_{FT(\nb)}$ in a very
efficient way,
as a sequence of
order($\nb^2$) elementary operations.
His decomposition will henceforth be called
the quantum Fast Fourier Transform (qFFT).
Another more
difficult desideratum
for a quantum compiler is that it
should
decompose $U_{FT(\nb)}$ into
a sequence of
order($\nb^2$) elementary operations.
Qubiter does this too.

Numerical evidence that Qubiter
can compile $H_\nb$ and
$U_{FT(\nb)}$ for
$\nb=1,2,3,4$ in this ideal
way was reported in Tuc99.
The goal of this paper is
to explain analytically
why Qubiter
behaves in this ideal way.
Qubiter
does not behave this way
because it is hardwired
to recognize
$H_\nb$ and
$U_{FT(\nb)}$.
Such a highly specialized
approach would be of limited scope.
Instead, the reason it behaves
this way is
because efficient expansions
of both
$H_\nb$ and
$U_{FT(\nb)}$ can both be viewed
as special cases of re-CSD,
and re-CSD is Qubiter's specialty.
This is a promising result.
It hints that
re-CSD is a door to
compiling efficiently
a large class of unitary
matrices that includes:
$H_\nb$,
$U_{FT(\nb)}$,
and an infinitude of
other matrices.

\section{Notation}
In this section,
we will introduce some notation
that is used throughout this paper.
For additional
information about notation,
the reader is referred to Ref.\cite{Paulinesia}.
Ref.\cite{Paulinesia} is a review paper
by the author of this paper
that uses the same notational
conventions as this paper.

For integers $a, b$ such that $a\leq b$, let
$Z_{a,b}=\{a, a+1, \ldots , b-1, b\}$.
We will often use
$\nb$ to denote the number of bits
in a quantum register, and
$\ns= 2^\nb$ to denote the
corresponding number of states.

First, let us introduce
the members of our cast of characters
that are 2d matrices.
The Pauli matrices are
\beq
\sigx =
\left[
\begin{array}{cc}
0&1\\
1&0
\end{array}
\right]
\;\;,\;\;
\sigy =
\left[
\begin{array}{cc}
0&-i\\
i&0
\end{array}
\right]
\;\;,\;\;
\sigz =
\left[
\begin{array}{cc}
1&0\\
0&-1
\end{array}
\right]
\;.
\eeq
This drama
will also feature
the 2d identity matrix
and the one-bit Hadamard matrix:

\beq
I =
\left[
\begin{array}{cc}
1&0\\
0&1
\end{array}
\right]
\;\;,\;\;
H =
\frac{1}{\sqrt{2}}
\left[
\begin{array}{cc}
1&1\\
1&-1
\end{array}
\right]
\;.
\eeq
Note that $H$ is
related to $e^{-i\frac{\pi}{4}\sigy}$,
a $\pi/2$ rotation about
the Y axis, as follows:

\beq
e^{-i\frac{\pi}{4}\sigy}=
\cos(\frac{\pi}{4}) -
i\sigy \sin(\frac{\pi}{4})=
\frac{1}{\sqrt{2}}
\left[
\begin{array}{cc}
1&-1\\
1&1
\end{array}
\right]=
H\sigz
\;.
\eeq

Of course, matrices of dimension greater than
2 will also make an appearance
 in this play. Some
will be built by using the tensor
product and direct sum of matrices.
In particular,
for any matrix $A$, we
can tensor-multiply or direct-sum several
copies of $A$.
For any positive integer $r$, let

\beq
A^{\otimes r}=
\underbrace{A\otimes A \otimes
\ldots \otimes A}_{r{\rm \;\;copies\;\;of\;\;}A}
\;,
\eeq
and

\beq
A^{\oplus r}=
\underbrace{A\oplus A \oplus
\ldots \oplus A}_{r{\rm \;\;copies\;\;of\;\;}A}
\;.
\eeq
For example,
$I^{\otimes r}\otimes A = A^{\oplus 2^r}$.
A fact that will be useful later on
is that for any two matrices $A,B$,
the transpose operation distributes over
$\oplus$ and $\otimes$:
$(A\oplus B)^T = A^T \oplus B^T$
and
$(A\otimes B)^T = A^T \otimes B^T$.

For any positive integer $r$,
let $D_r$:

\beq
D_r = (H\sigz)\otimes I^{\otimes r-1}=
\frac{1}{\sqrt{2}}
\left[
\begin{array}{cc}
I^{\otimes r-1} & -I^{\otimes r-1}\\
I^{\otimes r-1} & I^{\otimes r-1}
\end{array}
\right]
\;,
\label{eq-dr-mat}
\eeq
where
$I^{\otimes 0}=1$.
Note that
$D_r$
is $2^r$ dimensional.

\begin{figure}[h]
    \begin{center}
    \epsfig{file=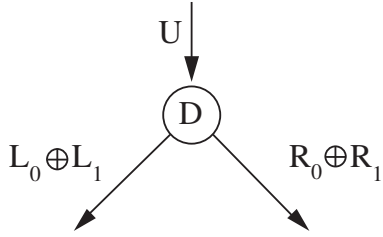, height=1.2in}
    \caption{Diagrammatic representation
    of Cosine-Sine Decomposition.}
    \label{fig-csd}
    \end{center}
\end{figure}

\begin{figure}[h]
    \begin{center}
    \epsfig{file=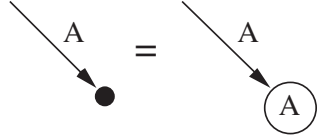, height=.7in}
    \caption{Definition of black-filled nodes.}
    \label{fig-black-node}
    \end{center}
\end{figure}

The main theme of this drama is
the recursive application of the CSD (re-CSD or
reap-CSD). The CSD (in the form used here)
is defined as follows. Given a
unitary matrix $U$ of even dimension $N$,
$U$ can be expressed as

\beq
U=(L_0\oplus L_1) D (R_0 \oplus R_1)
\;,
\label{eq-csd-formula}
\eeq
with

\beq
D =
\left[
\begin{array}{cc}
C & S\\-S & C
\end{array}
\right]
\;,
\eeq
and

\beq
C^2+S^2=1
\;,
\eeq
where $L_0, L_1, R_0, R_1$
are unitary matrices of dimension $N/2$,
and
where $C$ and $S$ are real diagonal matrices.
Eq.(\ref{eq-csd-formula})
is represented diagrammatically
in Fig.\ref{fig-csd}. Matrix $D$
is assigned to the node, matrix $U$
is assigned to the incoming arrow,
matrices
$L_0\oplus L_1$ and
$R_0\oplus R_1$ are each
assigned to an outgoing
arrow. When the CSD is used recursively,
then the various applications
of CSD can each be represent as in
Fig.\ref{fig-csd} and connected to form a
CSD binary tree.
We will refer to the
matrix assigned to the arrow entering
the root node of the CSD tree as
the {\bf initial matrix}, $U_{in}$, of the tree.
Fig.\ref{fig-black-node} shows
another convention for CSD trees
that will be used here. Namely, a
black-filled node will
represent a node that is
assigned the same matrix that
is assigned to the node's single incoming arrow.

\begin{figure}[h]
    \begin{center}
    \epsfig{file=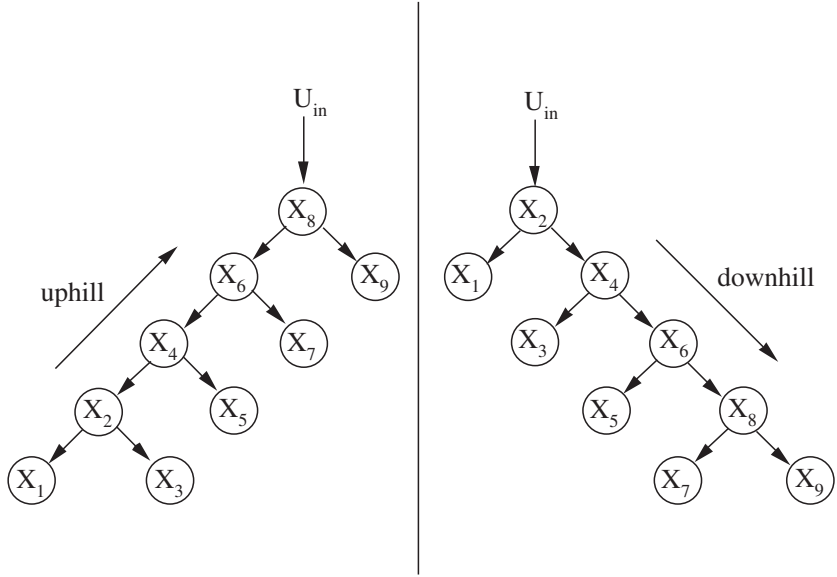, height=3in}
    \caption{CSD trees pruned so that
    they contain only leftmost
    or rightmost branches.}
    \label{fig-ud-hill}
    \end{center}
\end{figure}

In this paper, we are mostly concerned with
``degenerate" CSD trees that have been pruned so
that they contain only  the
leftmost or rightmost branches
(See Fig.\ref{fig-ud-hill}).
As discussed in
most books
about programming algorithms,
there
are several
algorithms for traversing all
the nodes of a tree.
The algorithm
that will be used in this paper
(also
used by Qubiter)
is one of the most common,
and is called the {\bf in-order
tree transversal strategy}.
In this strategy,
one visits (1) the left
sub-tree, (2)the root node,
(3) the right sub-tree,
in a recursive manner.
If one lists,
in accordance with the
in-order strategy,  the node labels
of either of
the two trees
in Fig.\ref{fig-ud-hill},
one obtains for both trees:

\beq
U_{in}=X_1 X_2 X_3\ldots X_9
\;.
\eeq
Henceforth we
will refer to the left and
right hand side trees
of Fig.\ref{fig-ud-hill}
as the {\bf uphill and downhill trees},
respectively.

\section{Hadamard Matrices}

In  this section, we
will consider
re-CSD with initial
matrix equal to
the $\nb$-bit Hadamard matrix $H^{\otimes \nb}$.
This problem is closely related
to the one considered in the
next section, re-CSD with initial matrix
equal to the $\nb$-bit
Discrete Fourier Transform matrix.
For simplicity, we will assume that
$\nb=4$. How to generalize our
results from $\nb=4$ to arbitrary
$\nb$ will be obvious.

\begin{figure}[h]
    \begin{center}
    \epsfig{file=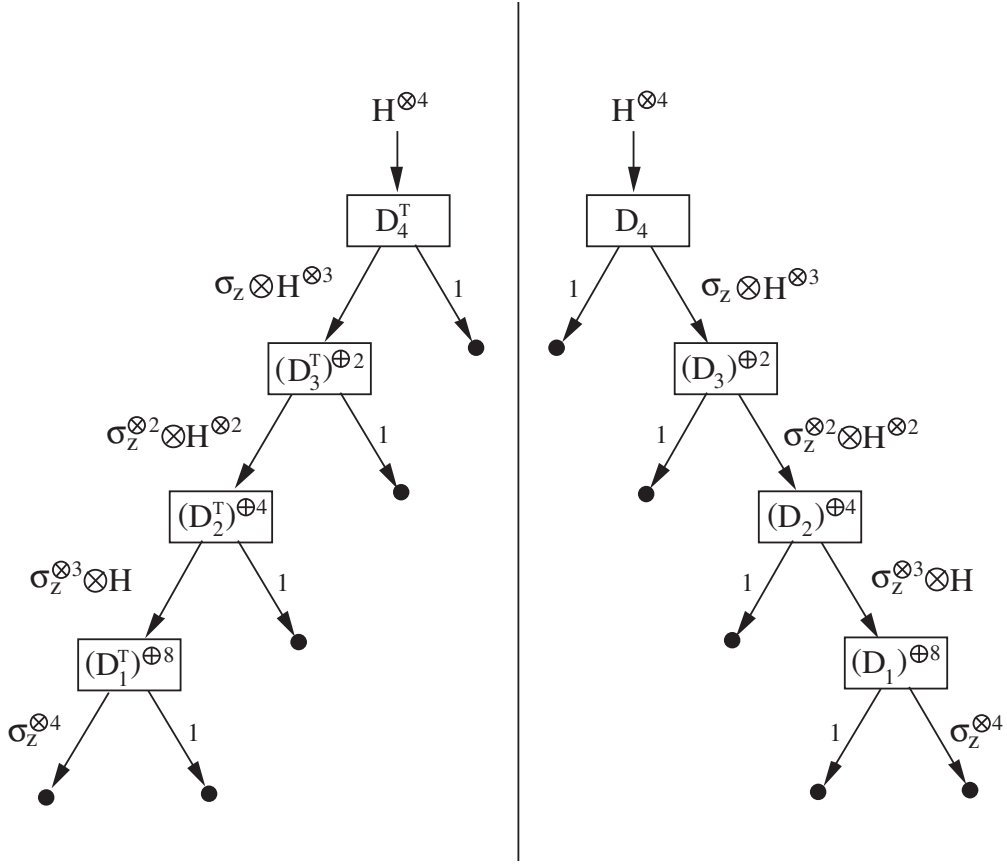, height=4.5in}
    \caption{re-CSD
    with initial matrix equal to
    the 4-bit Hadamard matrix.}
    \label{fig-had}
    \end{center}
\end{figure}

Fig.\ref{fig-had} shows the
CSD tree that is produced
by Qubiter when the
initial matrix is $H^{\otimes 4}$.
We will spend the
remainder of this
section
explaining Fig.\ref{fig-had}.
(As discussed in Tuc99,
the CSD
is not unique. Due to this non-uniqueness,
there are many possible CSD trees
that can be produced from compiling
the same initial matrix $H^{\otimes 4}$.
Fig.\ref{fig-had} is just one
of these possibilities.
Tuc99 discusses what choices
must be made in order to
steer Qubiter towards producing
this particular tree.)

We can express the initial matrix
$H^{\otimes 4}$
as a product of one-qubit
Hadamard matrices:
\beq
H^{\otimes 4}=
H(3) H(2) H(1) H(0)
\;.
\label{eq-had-3210}
\eeq
Eq.(\ref{eq-had-3210})
can be expressed recursively
as

\begin{subequations}
\label{eq-had-recursion-ops}
\beq
H^{\otimes 4}=
\Gamma(3210)
\;,
\eeq

\beq
\Gamma(3210)=
(H\sigz)(3)\sigz(3)\Gamma(210)
\;,
\eeq

\beq
\Gamma(210)=
(H\sigz)(2)\sigz(2)\Gamma(10)
\;,
\eeq

\beq
\Gamma(10)=
(H\sigz)(1)\sigz(1)\Gamma(0)
\;,
\eeq
and

\beq
\Gamma(0)=
(H\sigz)(0)\sigz(0)
\;.
\eeq
\end{subequations}

It is convenient to
translate the various bit-labelled operators
in Eqs.(\ref{eq-had-recursion-ops})
into matrices.
Define $\Gamma_r$ for
$r\in Z_{0, 4}$ by

\begin{subequations}
\label{eq-had-gamma-op-to-mat}
\beq
\Gamma(3210) = \Gamma_4
\;,
\eeq

\beq
\Gamma(210)= I \otimes \Gamma_3
\;,
\eeq

\beq
\Gamma(10) = I^{\otimes 2} \otimes \Gamma_2
\;,
\eeq

\beq
\Gamma(0) = I^{\otimes 3} \otimes \Gamma_1
\;,
\eeq
and

\beq
\Gamma_0 =1
\;.
\eeq
\end{subequations}
Note that for $r\in Z_{0,4}$,
$\Gamma_r$
is a matrix of dimension $2^r$.
In fact, $\Gamma_r = H^{\otimes r}$.
The bit-labelled operators
$(H\sigz)(\alpha)$
can be expressed in terms of the
$D_r$ matrices defined by Eq.(\ref{eq-dr-mat}):

\begin{subequations}
\label{eq-had-hsigz-op-to-mat}
\beq
(H\sigz)(3)=
D_4
\;,
\eeq

\beq
(H\sigz)(2)=
I\otimes D_3
\;,
\eeq

\beq
(H\sigz)(1)=
I^{\otimes 2}\otimes D_2
\;,
\eeq
and

\beq
(H\sigz)(0)=
I^{\otimes 3}\otimes D_1
\;
\eeq
\end{subequations}
Likewise, the bit-labelled operators
$\sigz(\alpha)$
can be expressed as matrices:

\begin{subequations}
\label{eq-had-sigz-op-to-mat}
\beq
\sigz(3)=
\left[
\begin{array}{cc}
I^{\otimes 3} & 0\\
0 & -I^{\otimes 3}
\end{array}
\right]
\;,
\eeq

\beq
\sigz(2)=
I\otimes
\left[
\begin{array}{cc}
I^{\otimes 2} & 0\\
0 & -I^{\otimes 2}
\end{array}
\right]
\;,
\eeq

\beq
\sigz(1)=
I^{\otimes 2}\otimes
\left[
\begin{array}{cc}
I & 0\\
0 & -I
\end{array}
\right]
\;,
\eeq
and

\beq
\sigz(0)=
I^{\otimes 3}\otimes \sigz
\;.
\eeq
\end{subequations}
After replacing bit-labelled
operators by their matrix equivalents
via
Eqs.(\ref{eq-had-gamma-op-to-mat}),
(\ref{eq-had-hsigz-op-to-mat})
and (\ref{eq-had-sigz-op-to-mat}),
the
recursion relation
defined by
Eqs.(\ref{eq-had-recursion-ops})
becomes simply:

\beq
\Gamma_{r+1}=
D_{r+1}
\left[
\begin{array}{cc}
\Gamma_r & 0\\
0 & -\Gamma_r
\end{array}
\right]
\;,
\label{eq-had-recursion-mats}
\eeq
for $r\in Z_{0,3}$.

The downhill tree of
Fig.\ref{fig-had}
was obtained using re-CSD in
combination with
Eq.(\ref{eq-had-recursion-mats})
and the following
identities:

\begin{subequations}
\beqa
\lefteqn{
\sigz \otimes H^{\otimes 3}
=}\nonumber\\&&
H^{\otimes 3} \otimes (-H^{\otimes 3})
\;,
\eeqa

\beqa
\lefteqn{
\sigz^{\otimes 2} \otimes H^{\otimes 2}
=}\nonumber\\&&
H^{\otimes 2} \otimes (-H^{\otimes 2})
\otimes
(-H^{\otimes 2}) \otimes H^{\otimes 2}
\;,
\eeqa

\beqa
\lefteqn{
\sigz^{\otimes 3} \otimes H
=}\nonumber\\&&
H \otimes (-H)\otimes(-H) \otimes H
\otimes
(-H) \otimes H \otimes H \otimes (-H)
\;,
\eeqa
and

\beqa
\lefteqn{
\sigz^{\otimes 4}
=}\nonumber\\&&
\sigz \otimes (-\sigz)
\otimes(-\sigz) \otimes \sigz
\otimes
(-\sigz) \otimes \sigz
\otimes \sigz \otimes (-\sigz)
\;.
\eeqa
\end{subequations}

Listing the node matrices
of the downhill tree
 given by Fig.\ref{fig-had}
(listing them
in the order visited by an
in-order tree transversal)
gives:

\beq
H^{\otimes 4} =
D_4
D^{\oplus 2}_3
D^{\oplus 4}_2
D^{\oplus 8}_1
\sigz^{\otimes 4}
\;.
\label{eq-had-fin-d}
\eeq
Using re-CSD, Qubiter
expresses $H^{\otimes 4}$
in the form given by
the right hand side of
Eq.(\ref{eq-had-fin-d}).
Then it
recognizes that: (1)the matrices
of the form
$D^{\oplus s}_r$
are one-qubit Y-axis rotations,
and (2) the matrix
$\sigz^{\otimes 4}$
is a
$\sigz$ matrix
applied separately to each qubit.

Note that Eq.(\ref{eq-had-3210})
listed the mutually commuting
operators $\{H(\alpha):\alpha\in Z_{0,3}\}$ in
 one of $4!$
equivalent orders.
If we take the transpose of
both sides of
Eq.(\ref{eq-had-3210}),
we reverse the order of the
$H(\alpha)$ operators:

\beq
H^{\otimes 4}=
H(0)H(1) H(2) H(3)
\;.
\label{eq-had-3210-transpose}
\eeq
In fact, we can take the transpose
of all
equations
 between and
 including Eqs.(\ref{eq-had-3210})
to
Eqs.(\ref{eq-had-fin-d}).
In particular, we get

\beq
\Gamma^T_{r+1}=
\left[
\begin{array}{cc}
\Gamma^T_r & 0\\
0 & -\Gamma^T_r
\end{array}
\right]
D^T_{r+1}
\;
\eeq
for $r\in Z_{0,3}$. Also,

\beq
H^{\otimes 4} =
\sigz^{\otimes 4}
(D^T_1)^{\oplus 8}
(D^T_2)^{\oplus 4}
(D^T_3)^{\oplus 2}
D^T_4
\;.
\label{eq-had-fin-d-tanspose}
\eeq
Eq.(\ref{eq-had-fin-d-tanspose})
also follows if we list the node
matrices
of the uphill tree
 given by Fig.\ref{fig-had}
(listing them
in the order visited by an
in-order tree transversal).

\section{Discrete Fourier Transform Matrices}

In this section, we will consider
re-CSD with initial
matrix equal to
the $\nb$-bit Discrete Fourier Transform matrix,
defined by $(U_{FT})_{x,y}
= \frac{1}{\sqrt{\ns}}e^{i\frac{2\pi xy}{\ns}}$,
where $x,y\in Z_{0,\ns-1}$.
For simplicity, we will assume that
$\nb=4$.

\begin{figure}[h]
    \begin{center}
    \epsfig{file=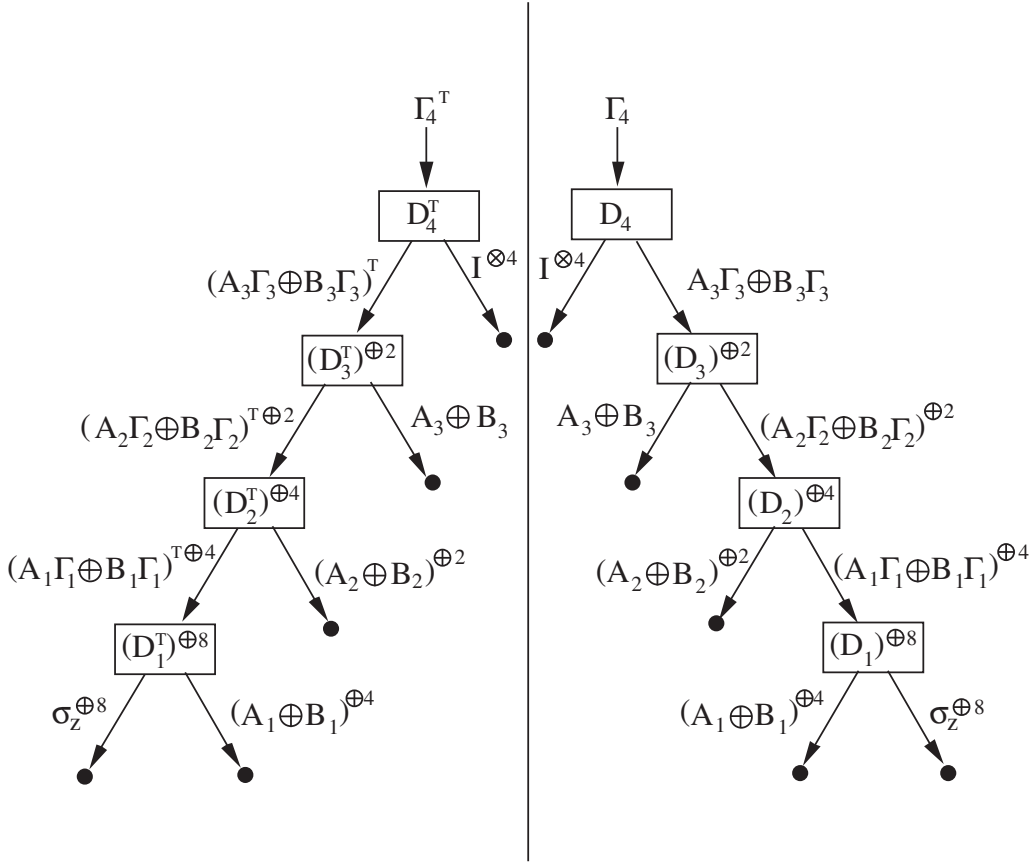, height=4.5in}
    \caption{re-CSD with initial matrix
    equal to the 4-bit
    Discrete Fourier Transform matrix.}
    \label{fig-fou}
    \end{center}
\end{figure}

Fig.\ref{fig-fou} shows the
CSD tree that is produced
by Qubiter when the
initial matrix is $U_{FT}$ for $\nb=4$.
We will spend the
remainder of this
section
explaining Fig.\ref{fig-fou}.

In Ref.\cite{Copper}, Coppersmith
showed how to express
$U_{FT}$
as a sequence of
order($\nb^2$) elementary operations.
We will call his decomposition
the quantum Fast Fourier Transform (qFFT).
For a pedagogical discussion
of the
qFFT circuit and
related matters, see Ref.\cite{Paulinesia}.
Ref.\cite{Paulinesia} has the virtue
that it uses
the same notation as this paper.

Define the root of unity $\omega$ and
the 2d matrix $\Omega$ by

\beq
\omega = \exp(i\frac{2\pi}{\ns})
\;,\;\;
\Omega=diag(1,\omega) = \omega^n
\;,
\label{eq-omega-defs}
\eeq
where $n$ is the number operator.

As in Ref.\cite{Paulinesia},
for any 2 distinct bits
$\alpha, \beta\in Z_{0, \nb-1}$,
let us define an operator $V(\alpha, \beta)$
by

\begin{equation}
\begin{array}{c}
\Qcircuit @C=1em @R=1.5em @!R{
&\dotgate\qwx[1]
&\qw
\\
&\dotgate
&\qw
}
\end{array}
=
V(\alpha, \beta)=
\exp[ i\pi
\frac{n(\alpha)n(\beta)}{2^{|\alpha-\beta|}}
]
\;.
\end{equation}

As in Ref.\cite{Paulinesia}, $R$
will denote the bit reversal matrix.
For $\nb=4$, it maps bits
$0\rarrow 3$,
$1\rarrow 2$,
$2\rarrow 1$, and
$3\rarrow 0$.
It can be expressed in terms of
exchange operators, which in turn
can be expressed in terms of CNOTs.

As shown in Ref.\cite{Paulinesia},
the qFFT for $\nb=4$
is

\beq
U_{FT}=
H(3)
V(3,2)V(3,1)V(3,0)
H(2)
V(2,1)V(2,0)
H(1)
V(1,0)
H(0)R
\;.
\label{eq-fou-3210}
\eeq
A diagrammatical way of saying the same thing is:

\begin{equation}
U_{FT}=
\begin{array}{c}
\Qcircuit @C=1em @R=.5em @!R{
&\qw
&\qw
&\qw
&\dotgate\qwx[3]
&\qw
&\qw
&\dotgate\qwx[2]
&\qw
&\dotgate\qwx[1]
&\gate{H}
&\multigate{3}{R}
\\
&\qw
&\qw
&\dotgate\qwx[2]
&\qw
&\qw
&\dotgate\qwx[1]
&\qw
&\gate{H}
&\dotgate
&\qw
&\ghost{R}
\\
&\qw
&\dotgate\qwx[1]
&\qw
&\qw
&\gate{H}
&\dotgate
&\dotgate
&\qw
&\qw
&\qw
&\ghost{R}
\\
&\gate{H}
&\dotgate
&\dotgate
&\dotgate
&\qw
&\qw
&\qw
&\qw
&\qw
&\qw
&\ghost{R}
}
\end{array}
\;.
\end{equation}
Note that Eq.(\ref{eq-fou-3210})
for $U_{FT}$  and Eq.(\ref{eq-had-3210})
for $H^{\otimes 4}$ are
very similar. They differ in that
the expression for $U_{FT}$
contains, in addition
to the one-bit Hadamard matrices,
 the bit reversal
matrix $R$, and
diagonal matrices inserted between the
one-bit Hadamard matrices.
It is convenient at this point
to lump together
the diagonal operators
that occur between the one-bit
Hadamard matrices. Define
diagonal operators
$\Delta(\cdot)$ by

\begin{subequations}
\label{eq-fou-Delta-ops}
\beq
\Delta(3210)=
\sigz(3)V(3,2)V(3,1)V(3,0)
\;,
\eeq

\beq
\Delta(210)=
\sigz(2)V(2,1)V(2,0)
\;,
\eeq

\beq
\Delta(10)=
\sigz(1)V(1,0)
\;,
\eeq
and

\beq
\Delta(0)=
\sigz(0)
\;.
\eeq
\end{subequations}

Next, let us consider re-CSD with
 the initial matrix
$U_{FT}R$. Eq.(\ref{eq-fou-3210})
can be expressed recursively as

\begin{subequations}
\label{eq-fou-recursion-ops}
\beq
U_{FT}R=
\Gamma(3210)
\;,
\eeq

\beq
\Gamma(3210)=
(H\sigz)(3)\Delta(3210)\Gamma(210)
\;,
\eeq

\beq
\Gamma(210)=
(H\sigz)(2)\Delta(210)\Gamma(10)
\;,
\eeq

\beq
\Gamma(10)=
(H\sigz)(1)\Delta(10)\Gamma(0)
\;,
\eeq
and

\beq
\Gamma(0)=
(H\sigz)(0)\Delta(0)
\;.
\eeq
\end{subequations}

It is convenient to
translate the various bit-labelled operators
in Eqs.(\ref{eq-fou-recursion-ops})
into matrices.
Define $\Gamma_r$ for
$r\in Z_{0, 4}$ by

\begin{subequations}
\label{eq-fou-gamma-op-to-mat}
\beq
\Gamma(3210) = \Gamma_4
\;,
\eeq

\beq
\Gamma(210)= I \otimes \Gamma_3
\;,
\eeq

\beq
\Gamma(10) = I^{\otimes 2} \otimes \Gamma_2
\;,
\eeq

\beq
\Gamma(0) = I^{\otimes 3} \otimes \Gamma_1
\;,
\eeq
and

\beq
\Gamma_0 =1
\;.
\eeq
\end{subequations}
Note that for $r\in Z_{0,4}$,
$\Gamma_r$
is a matrix of dimension $2^r$.
The bit-labelled operators
$(H\sigz)(\alpha)$
can be expressed in terms of the
$D_r$ matrices defined by Eq.(\ref{eq-dr-mat}):

\begin{subequations}
\label{eq-fou-hsigz-op-to-mat}
\beq
(H\sigz)(3)=
D_4
\;,
\eeq

\beq
(H\sigz)(2)=
I\otimes D_3
\;,
\eeq

\beq
(H\sigz)(1)=
I^{\otimes 2}\otimes D_2
\;,
\eeq
and

\beq
(H\sigz)(0)=
I^{\otimes 3}\otimes D_1
\;.
\eeq
\end{subequations}
The bit-labelled operators
$\Delta(\cdot)$
can be expressed in terms of the
$\Omega$ matrix defined by Eq.(\ref{eq-omega-defs}):

\begin{subequations}
\label{eq-fou-delta-op-to-mat}
\beqa
\Delta(3210)
&=&
\sigz(3)e^{i\pi[
\frac{n(2)}{2} +
\frac{n(1)}{4} +
\frac{n(0)}{8}]n(3)}\nonumber\\
&=&
\sigz(3)\omega^{i\pi[
4n(2) +
2n(1) +
n(0)]n(3)}\nonumber\\
&=&
\left[
\begin{array}{cc}
I^{\otimes 3} & 0\\
0 & -\Omega^4\otimes \Omega^2 \otimes \Omega
\end{array}
\right]\nonumber\\
&=&
A_3\oplus B_3
\;,
\eeqa

\beqa
\Delta(210)
&=&
\sigz(2)e^{i\pi[
\frac{n(1)}{2} +
\frac{n(0)}{4}]n(2)}\nonumber\\
&=&
\sigz(2)\omega^{i\pi[
4n(1) +
2n(0)]n(2)}\nonumber\\
&=&
I\otimes
\left[
\begin{array}{cc}
I^{\otimes 2} & 0\\
0 & -\Omega^4\otimes \Omega^2
\end{array}
\right]\nonumber\\
&=&I\otimes(A_2\oplus B_2)
\;,
\eeqa

\beqa
\Delta(10)
&=&
\sigz(1)e^{i\pi[
\frac{n(0)}{2}]n(1)}\nonumber\\
&=&
\sigz(1)\omega^{i\pi[
4n(0)]n(1)}\nonumber\\
&=&
I^{\otimes 2}\otimes
\left[
\begin{array}{cc}
I^{\otimes 2} & 0\\
0 & -\Omega^4
\end{array}
\right]\nonumber\\
&=&I^{\otimes 2}\otimes(A_1\oplus B_1)
\;,
\eeqa
and

\beqa
\Delta(0)&=&\sigz(0)\nonumber\\
&=& I^{\otimes 3} \otimes \sigz\nonumber\\
&=& I^{\otimes 3} \otimes (A_0 \oplus B_0)
\;.
\eeqa
\end{subequations}
The matrices $A_r$ and $B_r$,
where $r\in Z_{0,3}$,
are
first mentioned in
Eqs.(\ref{eq-fou-delta-op-to-mat}). They
are implicitly defined by those equations.
Note that for $r\in Z_{0,3}$,
$A_r$ and $B_r$
are diagonal unitary matrices of dimension $2^r$.
After replacing bit-labelled
operators by their matrix equivalents
via
Eqs.(\ref{eq-fou-gamma-op-to-mat}),
(\ref{eq-fou-hsigz-op-to-mat})
and (\ref{eq-fou-delta-op-to-mat}),
the
recursion relation
defined by
Eqs.(\ref{eq-fou-recursion-ops})
becomes simply:

\beq
\Gamma_{r+1}=
D_{r+1}
\left[
\begin{array}{cc}
A_r \Gamma_r & 0\\
0 & B_r\Gamma_r
\end{array}
\right]
\;,
\label{eq-fou-recursion-mats}
\eeq
for $r\in Z_{0,3}$.

The downhill tree of
Fig.\ref{fig-fou}
was obtained  using re-CSD in
combination with
Eq.(\ref{eq-fou-recursion-mats}).

Listing the node matrices
of the downhill tree
 given by Fig.\ref{fig-fou}
(listing them
in the order visited by an
in-order tree transversal)
gives:

\beq
U_{FT}R=
\Gamma_4=
D_4
(A_3\oplus B_3)
D_3^{\oplus 2}
(A_2\oplus B_2)^{\oplus 2}
D_2^{\oplus 4}
(A_1\oplus B_1)^{\oplus 4}
\sigz^{\oplus 8}
\;.
\label{eq-fou-fin-d}
\eeq
 Qubiter
first finds the necessary permutation $R$
using a strategy to be discussed
in the next section.
Then Qubiter uses
re-CSD to
express $U_{FT}R$
in the form given by
the right hand side of
Eq.(\ref{eq-fou-fin-d}).
Then it
recognizes that: (1) the matrices
of the form
$D^{\oplus s}_r$
are one-qubit Y-axis rotations
,  (2) the diagonal matrices
$\sigz^{\oplus 8}$
and $(A_r\oplus B_r)^{\oplus s}$
are expressible as products of doubly
controlled phase factors (such as $V$ ).

If we take the transpose of
both sides of
Eq.(\ref{eq-fou-3210}),
we reverse the order of all
the operators
on the right hand side:

\beq
U_{FT}=
R H(0)V(1,0)H(1)V(2,0)V(2,1)H(2)
V(3,0)V(3,1)V(3,2)
H(3)
\;.
\eeq
Diagrammatically,

\begin{equation}
U_{FT}=
\begin{array}{c}
\Qcircuit @C=1em @R=.5em @!R{
&\multigate{3}{R}
&\gate{H}
&\dotgate\qwx[1]
&\qw
&\dotgate\qwx[2]
&\qw
&\qw
&\dotgate\qwx[3]
&\qw
&\qw
&\qw
\\
&\ghost{R}
&\qw
&\dotgate
&\gate{H}
&\qw
&\dotgate\qwx[1]
&\qw
&\qw
&\dotgate\qwx[2]
&\qw
&\qw
\\
&\ghost{R}
&\qw
&\qw
&\qw
&\dotgate
&\dotgate
&\gate{H}
&\qw
&\qw
&\dotgate\qwx[1]
&\qw
\\
&\ghost{R}
&\qw
&\qw
&\qw
&\qw
&\qw
&\qw
&\dotgate
&\dotgate
&\dotgate
&\gate{H}
}
\end{array}
\;.
\end{equation}
In fact, we can take the transpose
of all
equations
 between and
 including Eqs.(\ref{eq-fou-3210})
to
Eqs.(\ref{eq-fou-fin-d}).
In particular, we get

\beq
\Gamma^T_{r+1}=
\left[
\begin{array}{cc}
 \Gamma^T_r A_r& 0\\
0 & \Gamma^T_r B_r
\end{array}
\right]
D^T_{r+1}
\;
\eeq
for $r\in Z_{0,3}$.
Also,

\beq
R U_{FT}=
\Gamma^T_4=
\sigz^{\oplus 8}
(A_1\oplus B_1)^{\oplus 4}
(D^T_2)^{\oplus 4}
(A_2\oplus B_2)^{\oplus 2}
(D^T_3)^{\oplus 2}
(A_3\oplus B_3)
D^T_4
\;.
\label{eq-fou-fin-d-tanspose}
\eeq
Eq.(\ref{eq-fou-fin-d-tanspose})
also follows if we list the node
matrices
of the uphill tree
 given by Fig.\ref{fig-fou}
(listing them
in the order visited by an
in-order tree transversal).

\section{Bit Permutations Before Each CSD}
In decomposing  $U_{FT}$
via re-CSD, we first
pre or post multiplied
$U_{FT}$ by the bit
reversal matrix $R$.
This example makes it clear
that in using re-CSD, before each application
of the CSD, it is helpful to
permute either the
rows or the columns
(or both) of the input
matrix $U$ in Fig.\ref{fig-csd}.
Only a certain type of permutation
 will work; it must
also be a bit
permutation,
so that it can be expressed
as a product of bit exchange operators,
which in turn can be expressed
as a product of CNOTs.
But what is a good
strategy
for choosing such a permutation?

Decompositions such as
the ones given here for $H^{\otimes 4}$
and $U_{FT}$ come from CSD trees
that contain only a single branch.
To promote the
growth of such degenerate trees,
we want to stunt the growth of either
the left matrices $L_0, L_1$ or
the right matrices $R_0, R_1$
in Fig.\ref{fig-csd}.
The left matrices
are ``stunted" if $L_0$
and $L_1$ are diagonal matrices.
Likewise, the right matrices are
``stunted" if $R_0$ and $R_1$
are diagonal.

The general CSD is

\beqa
\left[
\begin{array}{cc}
U_{00} & U_{01}\\
U_{10} & U_{11}
\end{array}
\right]
&=&
\left[
\begin{array}{cc}
L_0 & 0\\
0 & L_1
\end{array}
\right]
\left[
\begin{array}{cc}
C & S\\
-S & C
\end{array}
\right]
\left[
\begin{array}{cc}
R_0 & 0\\
0 & R_1
\end{array}
\right]\nonumber\\
&=&
\left[
\begin{array}{cc}
L_0CR_0 & L_0SR_1\\
L_1(-S)R_0 & L_1CR_1
\end{array}
\right]
\;.
\eeqa
When the left matrices $L_0, L_1$
are diagonal,
$U_{00}U^\dagger_{10}$
and $U_{01}U^\dagger_{11}$
are diagonal matrices.
Analogously, when the right matrices
$R_0, R_1$
are diagonal, $U_{00}U^\dagger_{01}$
and $U_{10}U^\dagger_{11}$
are diagonal matrices.
So probably a good strategy
for selecting a row and/or column permutation
to perform before each CSD is to
find a bit permutation that minimizes
the absolute value of
the off-diagonal elements of
$U_{00}U^\dagger_{10} \oplus U_{01}U^\dagger_{11}$
and/or
$U_{00}U^\dagger_{01} \oplus U_{10}U^\dagger_{11}$.
The idea is to coax the
growth of sparse trees
that have only a few branches
(rightmost and leftmost branches
might or might not be included in such a tree.)

\end{document}